# Improved Peak Cancellation for PAPR Reduction in OFDM Systems

Lilin Dan, Yue Xiao, Wei NI, and Shaoqian Li

**Abstract**

This letter presents an improved peak cancellation (PC) scheme for peak-to-average power ratio (PAPR) reduction in orthogonal frequency division multiplexing (OFDM) systems. The main idea is based on a serial peak cancellation (SPC) mode for alleviating the peak regrowth of the conventional schemes. Based on the SPC mode, two particular algorithms are developed with different tradeoff between PAPR and computational complexity. Simulation shows that the proposed scheme has a better tradeoff among PAPR, complexity and signal distortion than the conventional schemes.

## I. Introduction

Orthogonal frequency division multiplexing (OFDM) is an effective transmission technique for frequency-selective channels [1]. However, it suffers from high peak-to-average power ratio (PAPR) which results in a larger back-off and hence power loss for nonlinear amplifiers.

To alleviate this problem, many approaches have been proposed [2]. Among these approaches, the distortion-based technique is promising because it can efficiently reduce the signal peak by a preset peak cancellation function. For example, the well-known clipping approach [2] utilizes a simple pulse function to limit the peak value. However, the infinite frequency response of the pulse function will trigger out-of-band radiation. In this case, filtering is desired at the cost of peak regrowth [4]. For alleviating the peak

Lilin Dan,Yue Xiao and Shaoqian Li are with the National Key Lab of Communications, University of Electronic Science and Technology of China, E-mail: lilindan@hotmail.com.

Wei Ni is with Nakajima Laboratory, Department of Human Communication, The University of Electro-Communications,Tokyo, 182-8585, Japan.

regrowth, repeated clipping and filtering (RCF) [5] has been proposed, but its computational complexity is high. Recently, simplified clipping and filtering (SCF) [6] is proposed at the cost of PAPR degradation.

In addition to clipping-based approaches, peak cancellation (PC) is proposed to improve the original peak cancellation function. For example, in [7], the windowing function is considered as the peak cancellation function, so as to avoid the out-of-band radiation and reduce the complexity. However, for multiple peak values of OFDM signals, the side lobes of multiple widowing functions will interfere with each other in the time domain, which results in considerable peak regrowth and hence PAPR degradation.

In this letter, an improved peak cancellation scheme is proposed. The main idea is based on a serial peak cancellation (SPC) mode to mitigate the interferences among multiple windowing functions, so as to alleviate the peak regrowth of the conventional schemes. Based on SPC mode, two particular algorithms are developed with different tradeoff between PAPR and complexity. Simulation shows that the proposed scheme can reduce the PAPR more efficiently with lower complexity and lower distortion than current distortion-based schemes.

## II. Conventional Schemes

### A. OFDM and PAPR

For an OFDM system with $N$ subcarriers and an oversampling factor of $J$, the modulated signal $X_k, 0 \leq k \leq N-1$ is padded by $(J-1)N$ zeros, and then passed to an $JN$-point inverse fast Fourier transform(IFFT). The transmitted signal in the time domain is given by

$$s_n = \frac{1}{\sqrt{N}} \sum_{k=0}^{JN-1} X_k e^{j2\pi nk/JN}, n=0, 1, ..., JN-1. \qquad (1)$$

And the PAPR of OFDM signal is defined as

$$\text{PAPR} = \frac{\max_{0 \leq n \leq JN-1} |s_n|^2}{E\left[|s_n|^2\right]}, \qquad (2)$$

in which $E[\cdot]$ denotes the expectation value.

### B. Repeated Clipping and Filtering (RCF)

In RCF, the clipped signal is given by

$$s'_n = \begin{cases} s_n, \text{for } |s_n| \leq A \\ Ae^{j\phi_n}, \text{else} \end{cases}, \qquad (3)$$

in which $A$ is the threshold and $\phi_n$ the phase of $s_n$.

The clipped signal $s'_n$ is then transformed into the frequency domain by FFT. The out-of-band component is set to zero through filtering, and the filtered signal is transformed back into the time domain by IFFT.

The above processes are repeated until it reach the the maximum iteration number. However, since there are two FFTs per iteration, the computational complexity of RCF is considerably high.

*C. Simplified Clipping and Filtering (SCF)*

SCF is developed in [6] to simplify the processing of multiple iterations in RCF. In the frequency domain, the clipped signal $\hat{X}_k$ after $V$ iterations is approximated

$$\hat{X}_k = X_k - \bar{\beta} F_k, 0 \leq k \leq JN - 1, \quad (4)$$

in which $F_k$ denotes the in-band clipping distortion on the $k$th subcarrier, and $\bar{\beta}$ is an approximation factor given in reference [6]. Finally, $\hat{X}_k$ is transformed into the time domain by IFFT.

SCF approximates the processing of multiple iterations at the cost of PAPR degradation. Furthermore, it still needs three $JN$-point FFTs.

### III. IMPROVED PEAK CANCELATION SCHEME

*A. Serial Peak Cancelation*

The conventional peak cancellation (CPC) scheme replaces the original pulse function with a band-limited windowing function. The processed signal is expressed as

$$\tilde{s}_n = s_n + \sum_m c_{n,m} \, 0 \leq n \leq JN - 1, \quad (5)$$

in which $c_{n,m}$, $0 \leq n \leq JN - 1$, is the scaled widowing function (SWF) for the peak at position $m$

$$c_{n,m} = \alpha_m g_{n-m}. \quad (6)$$

Here, $\alpha_m$ is the scale factor

$$\alpha_m = -(1 - \frac{A}{|s_m|}) s_m, \quad (7)$$

$g_n$ is the windowing function, and $g_{n-m}$ is the cyclic shift of $g_n$ by $m$. In [7], $g_n$ is a sinc function

$$g_n = \sin(\pi n/N)/\pi n, \; -\frac{N_s}{2} + 1 \leq n \leq \frac{N_s}{2} \quad (8)$$

where $N_s$ is the length of the windowing function.

Equations (6) and (7) indicate that with given threshold and windowing function, the SWF is just a function of the input signal $s_m$. Therefore, all the SWFs are generated simultaneously and then overlapped in the time domain, which may cause interferes from the side lobs of $g_n$ resulting in increasing distortion and PAPR degradation.

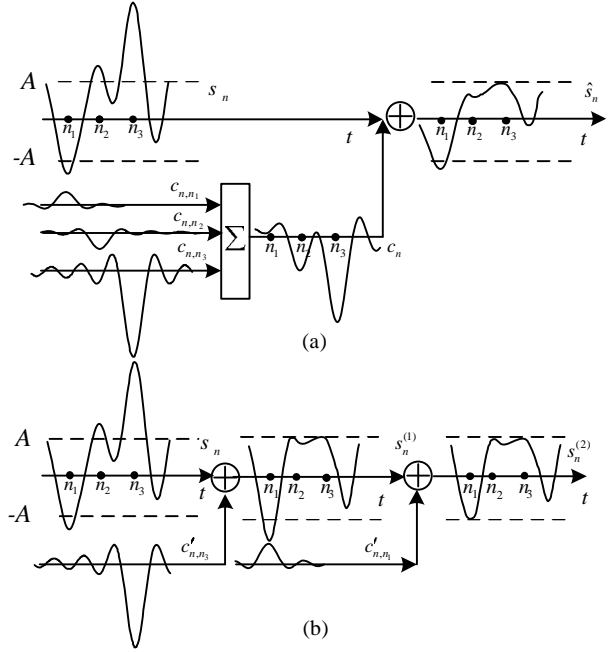

Fig. 1. Processing of different PC schemes ((a): CPC scheme; (b):the proposed SPC scheme)

Figure 1(a) gives an example of the signal processing by CPC. In the time domain, three SWFs, i.e., $c_{n,n_1}$, $c_{n,n_2}$ and $c_{n,n_3}$, are generated to limit the peaks at positions $n_1$, $n_2$ and $n_3$, respectively. As $c_{n,n_3}$ is negative to $c_{n,n_1}$ at $n_1$ while positive to $c_{n,n_2}$ at $n_2$, the interference from $c_{n,n_3}$ will aggravate the peak at $n_1$ while mitigate that at $n_2$. If the original OFDM signal is combined with three SWFs, the resultant signal suffers from peak regrowth at $n_1$ and non-necessary distortion at $n_2$.

To alleviate the above problems, a serial peak cancellation (SPC) mode is proposed, which processes the peaks in a serial manner as

$$s_n^{(i+1)} = s_n^{(i)} + c'_{n,m}, \tag{9}$$

$$c'_{n,m} = \alpha'_m g_{n-m}, \tag{10}$$

$$\alpha'_m = -(1 - \frac{A}{\left|s_m^{(i)}\right|})s_m^{(i)}. \tag{11}$$

Comparing Eq.(11) with Eq.(7), $\alpha'_m$ is no longer a function of $s_m$ but $s_m^{(i)}$, which is the updated OFDM signal after the $i$th peak cancelation. In this case, the scale factor $\alpha'_m$ is modified according to the interferences of previous processing. For example, in Fig.1(b), after adding $c'_{n,n_3}$ to the original OFDM signal, the peaks at $n_1$ and $n_2$ have been changed. Thus in the subsequent processing, SWF at $n_1$ is enhanced to mitigate the peak regrowth, while the cancellation at $n_2$ is canceled to avoid non-necessary distortion. In general, the proposed SPC mode can mitigate the interferences among SWFs, thus alleviate the peak regrowth and PAPR degradation.

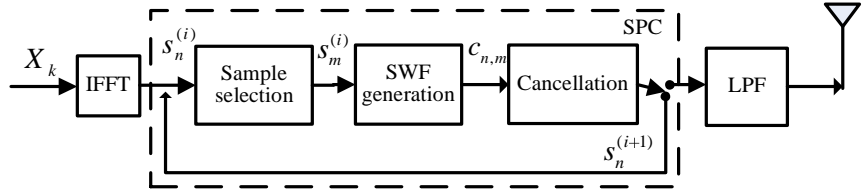

Fig. 2. Transmit blocks of the proposed SPC scheme.

The block diagram of the SPC-OFDM system is given in Fig.2. Firstly, the modulated signal $X_n$ is transformed into the time domain by IFFT. Then the peaks are detected and processed through the SPC block. Finally, the signal is passed to a low pass filter (LPF) to mitigate out-of-band radiation [4]. For SPC block, two particular algorithms are proposed.

*B. Improved Scheme (Algorithm 1)*

Algorithm I searches and limits the maximum peak iteratively until either no peak exceeds the threshold or it reaches the maximum iteration number $I_{max}$. The procedure is described as follows:

1) Let the counter $i$=0.
2) Detect the maximum peak with index

$$m = \arg\max_n \left\{ \left| s_n^{(i)} \right| \right\}. \quad (12)$$

3) If $\left| s_m^{(i)} \right| > A$, go to **4)**; else terminate processing.
4) Generate the SWF by Eqs.(10) and (11).
5) Limite the peak, and the processed signal is given by

$$s_n^{(i+1)} = \begin{cases} s_n^{(i)} + c'_{n,m}, & -\frac{N_s}{2}+m+1 \leq n \leq \frac{N_s}{2}+m \\ s_n^{(i)}, \text{else} \end{cases}. \quad (13)$$

6) $i$=$i$+1; if $i < I_{max}$ go to **2)** ; else terminate processing.

If the average number of iterations for each OFDM symbol is defined as $\bar{I}$ that $\bar{I} \leq I_{\max}$, the proposed algorithm needs one $JN$-point IFFT, $\bar{I}$ times of maximum searching and $\bar{I}$ times of peak cancellation. More particularly, a $JN$-point IFFT requires $\frac{JN}{2} \cdot \log_2 JN$ complex multiplications and $JN \cdot \log_2 JN$ complex additions. For the maximum searching in Eq.(12), in the first iteration it requires $JN$ complex multiplications for calculating $|s_n|$ and ($JN$-1) real comparisons; for the rest of the iterations, it only requires $N_s$ complex multiplications for calculating $\left|s_n^{i+1}\right|$ and ($JN$-1) comparisons. For the peak cancelation in Eqs.(10), (11) and (13): Eq.(10) requires $\frac{N_s}{2}+1$ real-complex multiplications; Eq.(11) requires one real-complex multiplication, one real multiplication and one real addition; and Eq.(13) requires $N_s$ complex additions. Recall that a complex addition is equivalent to two real additions; a complex multiplication to four real multiplications and two real additions; a real-complex multiplication to two real multiplications, the complexity of Algorithm I by real multiplications $C_{mul}$, real additions $C_{add}$ and real comparison $C_{comp}$ is

$$\begin{aligned} C_{mul} &= 2JN \cdot \log_2 JN + 4(JN - N_s) + 5\bar{I}(N_s+1), \\ C_{add} &= 3JN \cdot \log_2 JN + 2(JN - N_s) + \bar{I}(4N_s+1), \\ C_{comp} &= JN\bar{I}. \end{aligned} \quad (14)$$

*C. Improved Scheme (Algorithm 2)*

To make a better tradeoff between complexity and PAPR, a revised processes, Algorithm 2, is proposed. Instead of maximum searching, Algorithm 2 randomly selects a single sample from $JN$ samples without repetition. Thus a total of $JN$ iterations are required. During each iteration, the selected sample is processed by peak cancelation if its amplitude exceeds the threshold. The details of Algorithm 2 are described as follows:

1) Let the counter $i$=0.
2) Randomly select the index number $m$ from $\{0, \ldots, JN-1\}$ without repetition among iterations.
3) If $\left|s_m^{(i)}\right| > A$, go to **4)**; else go to **5)**.
4) Generate the scaled windowing function by Eqs.(10) and (11), and calculate $s_n^{(i+1)}$ by Eq.(13).
5) $i=i+1$; if $i < JN$, go to **2)**; else, terminate processing.

Through the above processing, the computational complexity for exhaustive searching is saved while the PAPR performance will be moderately degraded.

Define $\bar{P}$ as the average number of samples exceeding the threshold in one OFDM symbol, and assuming $X_n$ a random signal with zero mean and variance $\sigma^2$. Based on the central limit theorem, for an OFDM system with large number of subcarriers, the magnitude of $s_n$ approximates a Rayleigh process, thus $\bar{P}$ can be approximated by

$$\bar{P} \leq JN \cdot \Pr\{\left|s_n^{(i)}\right|^2 \geq A\} = JN \cdot e^{-A^2/\sigma^2} \ . \tag{15}$$

Algorithm 2 consists of one $JN$-point IFFT, $JN$ peak calculations ($JN$ amplitude calculations and $JN$ comparisons), and $\bar{P}$ peak cancellations. Thus the complexity of Algorithm 2 is given by

$$\begin{aligned} C_{mul} &= 2JN \cdot \log_2 JN + 4JN + \bar{P}(N_s+5), \\ C_{add} &= 3JN \cdot \log_2 JN + 2JN + \bar{P}(2N_s+1), \\ C_{comp} &= JN. \end{aligned} \tag{16}$$

## IV. NUMERICAL RESULTS

The performance of the proposed scheme is evaluated via simulation. The OFDM system considered is with 1024 subcarriers, $J$=4 and 16QAM modulation. The performance of the conventional schemes such as RCF [5], SCF [6] and CPC [7] are also simulated. For a fair comparison, the signal after PAPR reduction process is passed to a low-pass filter [4], and then normalized before PAPR calculation. In Algorithm 1, $I_{max}$ is selected larger than $\bar{P}$ to ensure that all the over-threshold peaks have been processed. In SCF, a sufficiently large $V$, i.e., 100, is set to achieve the up-limited PAPR reduction. In RCF, the selected $V$ is a tradeoff according to system desire, such as PAPR level, distortion toleration and computational complexity. In distortion-based schemes, there are three key factors for performance evaluation: PAPR reduction, computational complexity, and signal to distortion ratio (SDR). To make a simple comparison, the method of fixing certain factors and evaluating the rest are used. Since the proposed scheme has only one IFFT, its complexity is always lower than multi-IFFT schemes, i.e., RCF (($2V$-1) IFFTs) and SCF (3 IFFTs). Therefore, the following comparisons are mainly based on fixed PAPR or fixed SDR level.

Table 1 lists the PAPR, SDR and complexity of different distortion-based scheme with different thresholds. The comparison is mainly for the results with $A$=6dB to follow the literatures of current schemes [5] [6]. It shows that as the iteration of RCF increases, the SDR degradation and the PAPR gain decrease while the complexity increases. And with a comparable SDR to other schemes (around 24dB with $A$=6dB), the PAPR performance with $V$=5 is shown to be relatively efficient compared to other iterative cases. As shown in Table 1, the proposed scheme has much better PAPR and SDR performance than CPC, although the complexity is slightly increased. Furthermore, Algorithm 1 achieves 0.31dB PAPR reduction with only 23% complexity to a 5-iteration RCF, while it achieves 0.58dB PAPR reduction with only 84% complexity to SCF. And Algorithm 2 has the lowest complexity among all the schemes, whose complexity is only 12% of 5-iteration RCF and 38% of SCF. Considering the PAPR performance, Algorithm 2 is shown to outperform the other simulated schemes in terms of tradeoff among complexity, PAPR and BER performance.

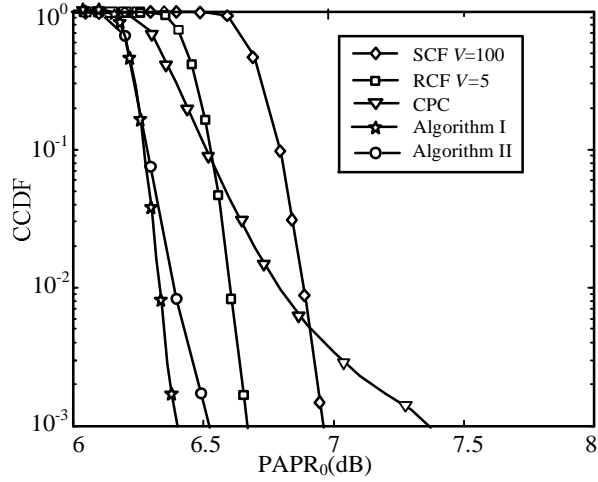

Fig. 3. PAPR performance of the proposed and current distortion-based schemes (*A*=6dB).

The performance of different schemes with comparable SDR is further illustrated in Fig.3 and Fig.4. As shown in Fig.3, with fixed SDR, the proposed scheme achieves better PAPR performance than current schemes in terms of the complementary cumulative distribution function (CCDF). From Fig.3, it is shown that CPC has the worst PAPR performance, especially at high $PAPR_0$. The reason lies in that the interferences among different SWFs will cause peak regrowth, which becomes dominating for high $PAPR_0$. On the other hand, since the improved scheme mitigates the interferences, it can alleviate the peak regrowth and achieve efficient PAPR reduction as shown in Fig.3. From Fig.4, it is shown that over AWGN channel, the schemes with comparable SDR,i.e., the proposed algorithms, SCF and 5-iteration RCF, has similar BER performance while the case with worst SDR, i.e., CPC, exhibits an error floor at BER of $10^{-2}$. When the SNR is high, the signal distortion becomes dominating, which influences the system performance. Since the improved scheme mitigates the non-necessary distortion, the system performance is improved even at high SNR.

Furthermore, similar comparison can be provided with other threshold, e.g. *A*=4dB as shown in Table 1 and Fig.5, which also shows that the proposed scheme achieves better performance than current schemes.

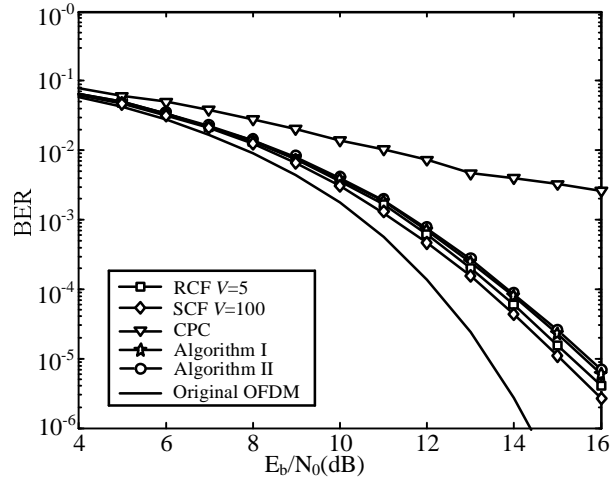

Fig. 4. BER performance of the proposed and current distortion-based schemes (*A*=6dB).

TABLE I

COMPARISONS OF THE PROPOSED AND CURRENT SCHEMES VIA DIFFERENT THRESHOLDS.(*PAPR:PAPR$_0$ AT $10^{-3}$ CCDF)

| Threshold | Merit | SPC | | CPC | SCF | RCF | | |
|---|---|---|---|---|---|---|---|---|
| | | Algorithm 1 | Algorithm 2 | | *V*=100 | *V*=1 | *V*=3 | *V*=5 |
| *A*=6dB | Multiplications | 120568 | 164788 | 122715 | 313573 | 196608 | 589824 | 967802 |
| | Addition | 166904 | 195548 | 171015 | 452684 | 197508 | 787332 | 1368446 |
| | *PAPR (dB) | 6.35 | 6.52 | 7.38 | 6.73 | 8.57 | 7.12 | 6.66 |
| | SDR (dB) | 24.33 | 24.16 | 17.74 | 25.57 | 29.86 | 25.64 | 24.58 |
| *A*=4dB | Multiplications | 264273 | 138208 | 149573 | 314341 | 115684 | 543662 | 971639 |
| | Addition | 274545 | 200672 | 222429 | 452940 | 155981 | 762852 | 1368446 |
| | *PAPR(dB) | 4.97 | 5.40 | 11.21 | 6.06 | 7.44 | 5.52 | 4.97 |
| | SDR (dB) | 17.56 | 17.28 | 8.74 | 19.32 | 22.17 | 17.98 | 16.81 |

For example, Algorithm 1 and Algorithm 2 achieve comparable PAPR to a 5-iteration and 3-iteration RCF with around 23% and 25% complexity, respectively. Compared with SCF, Algorithm 1and Algorithm 2 outperform 1.09dB and 0.66dB PAPR reduction with 70% and 44% complexity of SCF, respectively.

Here, we further investigate the SDR and complexity performance with a fixed PAPR. As shown in Fig.6, with various combinations of *A* and *V* the simulated RCF and SCF provide comparable PAPR to the proposed scheme, i.e., around 6.4dB at 0.1% CCDF. With the fixed PAPR, the proposed scheme has a better tradeoff between SDR and complexity to clipping-based schemes as shown in Table II. For example, the SDR of Algorithm 2 is 5dB higher than 1-iteration RCF ($A$=2.5dB) with comparable complexity, and the complexity of Algorithm 2 is only 15% to 5-iteration RCF ($A$=5.5dB) with comparable SDR. Furthermore, the SDR of Algorithm 2 is 0.4dB higher than SCF while the complexity is only 40% to SCF.

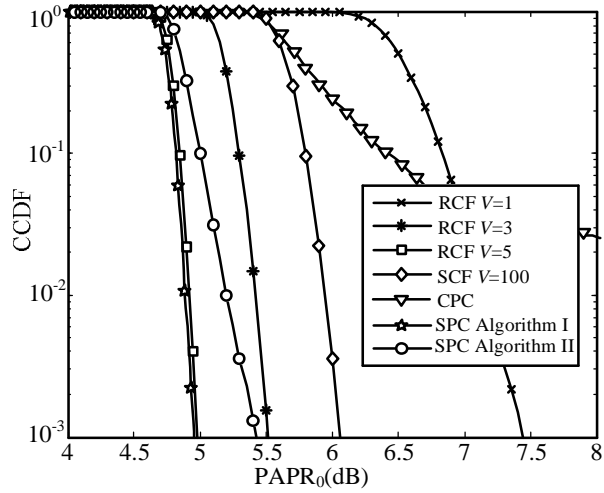

Fig. 5. PAPR performance of the proposed and current distortion-based schemes ($A$=4dB).

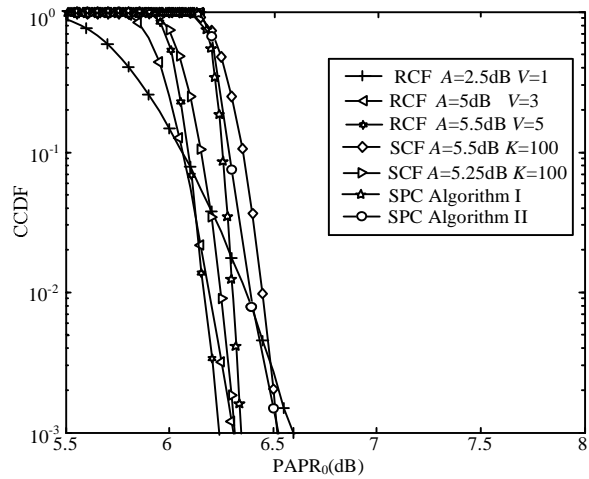

Fig. 6. PAPR performance of different schemes with different combination of $A$ and $V$.

TABLE II

Comparisons of the proposed and current schemes with comparable PAPR.(*PAPR: $PAPR_0$ at $10^{-3}$ CCDF)

|     |                      | *PAPR(dB) | SDR (dB) | IFFTs |
|-----|----------------------|-----------|----------|-------|
| RCF | $A$=2.50dB $V$=1     | 6.6       | 18.24    | 1     |
|     | $A$=5.00dB $V$=3     | 6.31      | 21.56    | 5     |
|     | $A$=5.50dB $V$=5     | 6.24      | 24.06    | 9     |
| SCF | $A$=5.25dB $V$=100   | 6.31      | 22.89    | 3     |
|     | $A$=5.50dB $V$=100   | 6.51      | 23.76    | 3     |
| SPC | Algorithm 1          | 6.35      | 24.33    | 1     |
|     | Algorithm 2          | 6.52      | 24.16    | 1     |

## V. Conclusions

An improved peak cancellation scheme is proposed for PAPR reduction in OFDM systems. It aims to mitigate the interference through optimized processing order in a serial manner. Furthermore, two particular SPC algorithms are described with different complexity. The comprehensive comparisons show that the proposed scheme outperforms the current distortion-based schemes, in terms of tradeoff among PAPR, computational complexity and signal distortion.